\documentclass[11pt]{article}

\usepackage[pdftex]{graphicx}
\usepackage[numbers]{natbib}
\usepackage{amssymb,amsmath}
\usepackage[a4paper,left=2cm,right=2cm,top=3cm,bottom=2.5cm]{geometry}

\def\nbOne{{\mathchoice {\rm 1\mskip-4mu l} {\rm 1\mskip-4mu l}
{\rm 1\mskip-4.5mu l} {\rm 1\mskip-5mu l}}}
\begin{document}

\title{Introduction to stellar coronagraphy}

\author{A. Ferrari\\
Fizeau, Universit\'e de Nice Sophia-Antipolis,  CNRS,
Observatoire de la C\^ote d'Azur \\ Parc Valrose, 06108 Nice, France
\and
R. Soummer\thanks{R. Soummer is supported by a AMNH Kalbfleisch Research Fellowship.} \\
American Museum of Natural History, \\ 79th St. at Central Park West, New York, NY 10024, USA 
\and
C. Aime\\
Fizeau, Universit\'e de Nice Sophia-Antipolis,  CNRS,
Observatoire de la C\^ote d'Azur\\Parc Valrose, 06108 Nice, France
}

\maketitle
\begin{abstract}
This paper gives a simple and original presentation of various coronagraphs inherited from the Lyot coronagraph. 
We first present the Lyot and Roddier phase mask coronagraphs and study their properties as a function of the focal mask size.
We show that the Roddier phase mask can be used to produce an apodization for the star.
Optimal coronagraphy can be obtained from two main approaches, using prolate spheroidal pupil apodization and a finite-size focal mask, or using a clear aperture and an infinite mask of variable transmission.

\end{abstract}

\section{Introduction}

Direct imaging of faint companions or planets around a bright star is a very difficult task, where the contrast ratio and the angular separation are the observable parameters. The problem consists of detecting a faint source over a bright and noisy background, mainly due to the diffracted stellar light. High contrast ratios and small angular separations correspond to the most difficult case. Typically, for giant exoplanets, contrast ratios of about $10^{-7}$ are expected in the near infrared (J;H;K bands), based on models for relatively young objects of about 100 Myr \citep{chabrier-2000-38,baraffe-2003-402,burrows-2004-609}. According to these models, older objects would be an order of magnitude fainter. Terrestrial planets are much fainter than giant planets, about 3 to 4 orders magnitudes fainter depending on the wavelength range.

The aim of this article is to give a short review of basic concepts and techniques
 used in focal plane mask coronagraphy for exoplanets detection. The paper will 
focus on derivation of a general formalism which allows to gain deeper insight in the behaviour of the principal coronagraphic techniques. It will not consider important problems related for example to the effects of adaptive optics residual speckles, slow-varying  speckles  caused by  mechanical or thermal deformations, telescope central obstruction, chromaticity. 

\section{Coronagraph general formalism}

A common  formalism can be used for    Lyot and Roddier \& Roddier 
coronagraphs \citep{lyot39,rodd97} in their classical or apodized
version \citep{artaa01b, aim03,soummer2005APJ}, four quadrants coronagraph (4QC) \citep{roua00},
band-limited mask coronagraph  \citep{kuch02} and shaped pupil coronagraph
\citep{kasd03}.

\begin{figure}
\centerline{\includegraphics[width=0.5\textwidth]{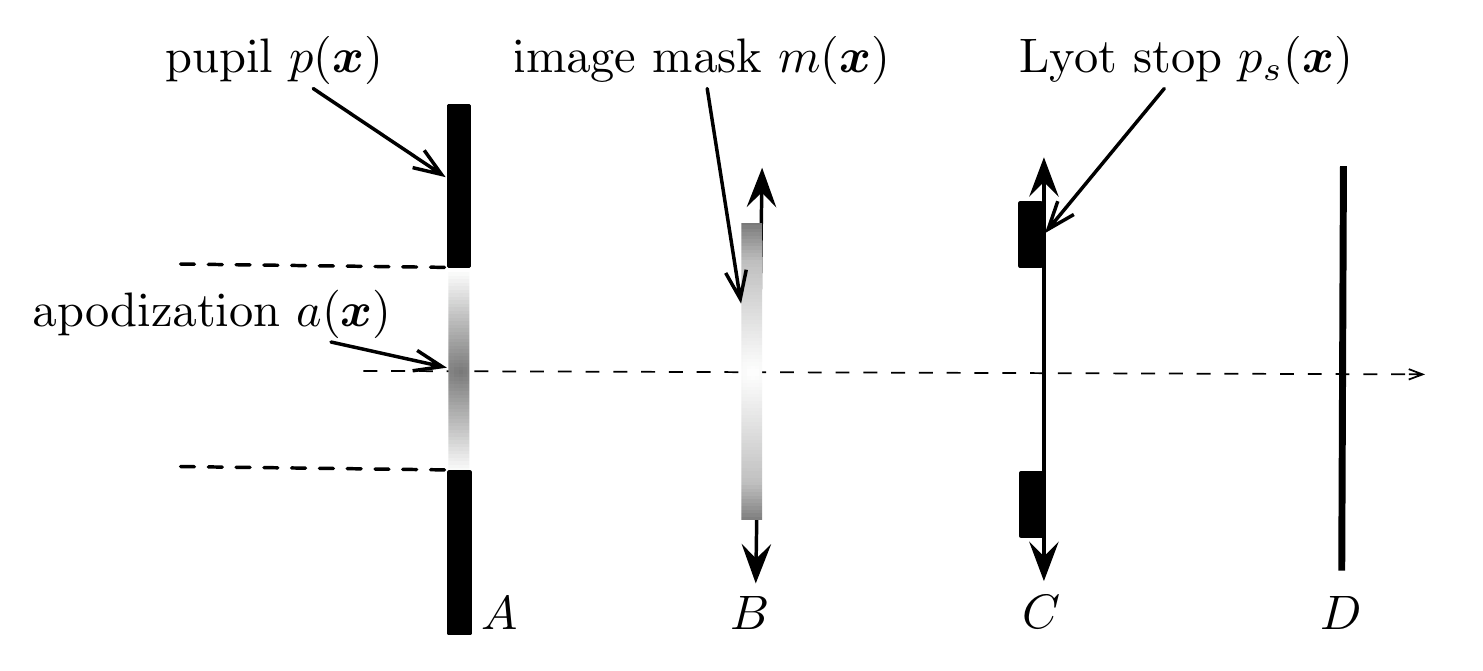}}
\caption{Schematic illustration of  Lyot, Roddier \& Roddier, 4QC and
band-limited mask coronagraph.\label{fig:setup}}
\end{figure}

We will follow the notations of \citep{artaa01b}.
The successive planes  of the coronagraph are denoted by $A$, $B$, $C$ and $D$. 
$A$ is the entrance aperture, $B$ denotes the focal plane
with the image mask, $C$ is the image  of the aperture with
the Lyot stop and $D$  is the image in the focal plane after the coronagraph. 
The general setup is illustrated in figure \ref{fig:setup}. The  position vector in each plane of the coronagraph will be noted in bold.

\begin{itemize}
\item We denote by $p(\vec{x})$ the telescope pupil and $a(\vec{x})$ its possible transmission 
if a pupil apodization is used. We assume that $\vec{x}$  in $A$ is in 
units of wavelength.

\item The mask transmission is  $t(\vec{x})$.  We will consider without loss of generality
 that $t(\vec{x})  =1-m(\vec{x})$. The units for the coordinates in $B$ and $D$ are angles on the sky.  
 
 \item The Lyot stop, denoted as $p_s(\vec{x})$ 
 will clearly acts as a low pass filter and consequently its size will be 
 chosen at most equal to the size of  the entrance pupil so that
  $p_s(\vec{x}) p(\vec{x})= p_s(\vec{x})$. 
  
 \end{itemize}
We denote as $\mathcal{P}$, $\mathcal{S}$ and $\mathcal{M}$ the areas
defined by the pupil, the stop and if pertinent the mask.
We will make the usual approximations of paraxial optics.
Moreover we neglect the quadratic phase terms associated with the propagation 
of the waves or assume that the optical layout is properly designed to cancel it~\citep{aime05}.
The constant propagation terms between the successive planes will be omitted.
In this case the coronagraph can be  easily  described using classical Fourier optics:
a Fourier transform exists between each of the four planes. 

The wavefront complex amplitude just before the pupil plane is $\Psi(\vec{x})$.
The expression of  the complex amplitude in the successive planes $A$, $B$, $C$ and $D$ are: 
\begin{eqnarray}
\Psi_A(\vec{x})&=&\Psi(\vec{x}) p(\vec{x})a(\vec{x}) \\
\Psi_B(\vec{x})&=& t(\vec{x})\hat{\Psi}_A(\vec{x})
= \hat{\Psi}_A(\vec{x}) - m(\vec{x})\hat{\Psi}_A(\vec{x}) \label{planBcart}\\
\Psi_C(\vec{x})&=& \hat{\Psi}_B(\vec{x})p_s(\vec{x})   
= \big(\Psi_A(\vec{x}) -
\Psi_A(\vec{x}) \ast \hat{m}(\vec{x}) \big) p_s(\vec{x}) \label{planCcart} \\
\Psi_D(\vec{x})&=& \hat{\Psi}_C(\vec{x})
=\big(\hat{ \Psi}_A(\vec{x}) - \hat{\Psi}_A(\vec{x}) m(x)\big)\ast  \hat{p}_s(\vec{x}) \label{planDcart}
\end{eqnarray}
where $\hat{g}$ is the Fourier transform of $g$ and $\ast$ denotes convolution.

The effect of the coronagraph clearly appears in equation (\ref{planCcart}). The first term is
the direct wave diaphragmed by the Lyot stop. The second term corresponds to
the wave diffracted by the mask for which the light diffracted outside the aperture in $C$ has been removed.
A coronagraph correctly designed for exoplanets imaging can operate one of these two techniques:
\begin{enumerate}
\item cancel the on-axis star without altering an off-axis source: 
for the star the two terms in (\ref{planCcart}) must interfere destructively whereas for the planet
 the  second term of (\ref{planCcart})  must cancel. In this case the response of the coronograph is no longer invariant by translation.
\item concentrate the on-axis star light reducing the off-axis diffracted light.
This approach corresponds to the apodization techniques \citep{aime05}.
\end{enumerate}
These two techniques can be achieved by a proper choice of the apodization $a(\vec{x})$
and the transmission $t(\vec{x})$. This article will focus on the first solution. For a general overview
of pure apodization techniques see \citep{aime05} and included references.

 Section 2 will present the main solutions 
to this problem that have been derived in the litterature. First the ``historical'' Lyot coronagraph and the Roddier coronagraph are presented. Whereas the first attempts to
cancel the star light, it appears that the second one can also operate by an apodization of the star light. Whereas these coronagraph can provide acceptable results when the dynamic is not too high
their performances appear to be insufficient  for the detection of earth-like planets.
Section 3 addresses the problem of
optimal coronagraphy which try to maximize a criterion quantifying the star light rejection. In this context  a complete star light rejection can be
obtained using an infinite size mask (4QC and band-limited mask) or a finite size 
$\pi$ phased mask (Roddier coronagraph)
 with a properly apodized entrance pupil. In the case of a Lyot coronagraph, an apodized aperture
 allows to maximize the star light rejection for  a given mask size. 
Through   all this presentation we will try to be, when possible,
as general as possible with respect to the geometry of the system without
specializing for example on a specific  pupil shape.

\section{Lyot and Roddier coronagraphs}

The first solution was proposed by Lyot  \citep{lyot39} and consists in canceling
 the major contribution of the star energy located in the telescope point spread function (PSF)
inside a disk of radius $r_m$.
This is simply achieved setting  for $t(\vec{x})$ an opaque mask at the center of the
image plane $B$.

The results will be presented for a circular aperture of radius $R$ and without central obstruction. 
The use of polar coordinates will be preferred. Under this asumption $p(\vec{x}) = \Pi(r/R)$, $m(\vec{x}) = \Pi(r/r_m)$
and $p_s(\vec{x})= \Pi(r/r_s)$ where $\Pi(r) = 1$ for $r\in [0,1)$  and $0$ if $r\geq 1$ and for example:
$\hat{p}(\vec{x})=Rr^{-1}J_1(2\pi R r)$.
Note that as long as $\Psi_A(\vec{x})$,  $a(\vec{x})$ and $m(\vec{x})$ are
radial functions, the complex amplitudes at the different stages of the coronagraph will exhibit
the same symmetry.

\begin{figure}[ht]
\centerline{\includegraphics[width=.8\textwidth]{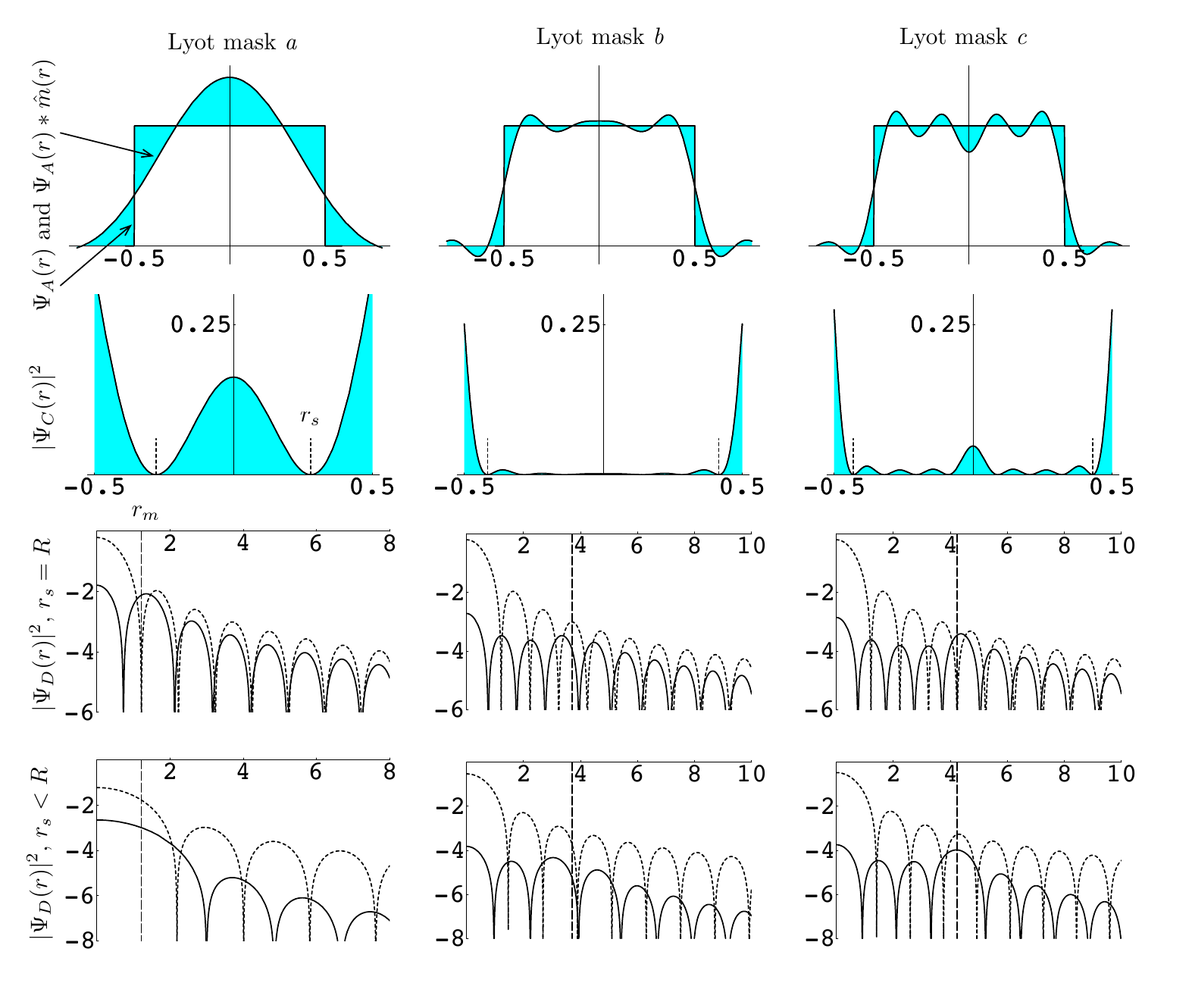}}
\caption{Illustration of Lyot coronagraphy for circular aperture (radial cuts) and an on-axis  point source.
The first raw represents the two complex amplitudes 
that should interfere destructively in $C$.
The second row gives the intensity in $C$ and the third and fourth
raw the intensity in $D$.
In the two last rows, the continuous line represents the intensity in $D$ for a coronagraphed
point source and the dashed line the PSF associated to a pupil of radius $r_s$.
The vertical dashed line in the last row indicates the radius of the mask. \label{fig:lyot}}
\end{figure}

\begin{figure}[ht]
\centerline{\includegraphics[width=.4\textwidth]{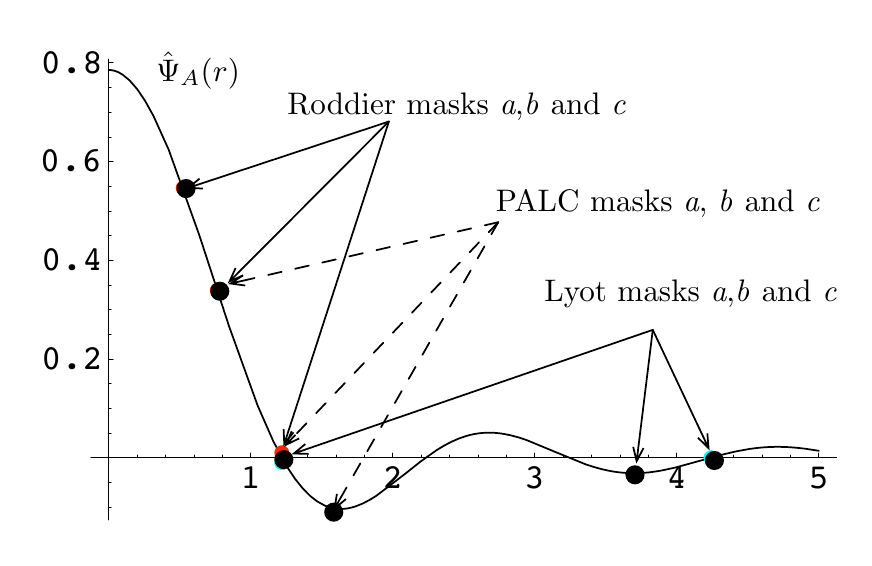}\includegraphics[width=.45\textwidth]{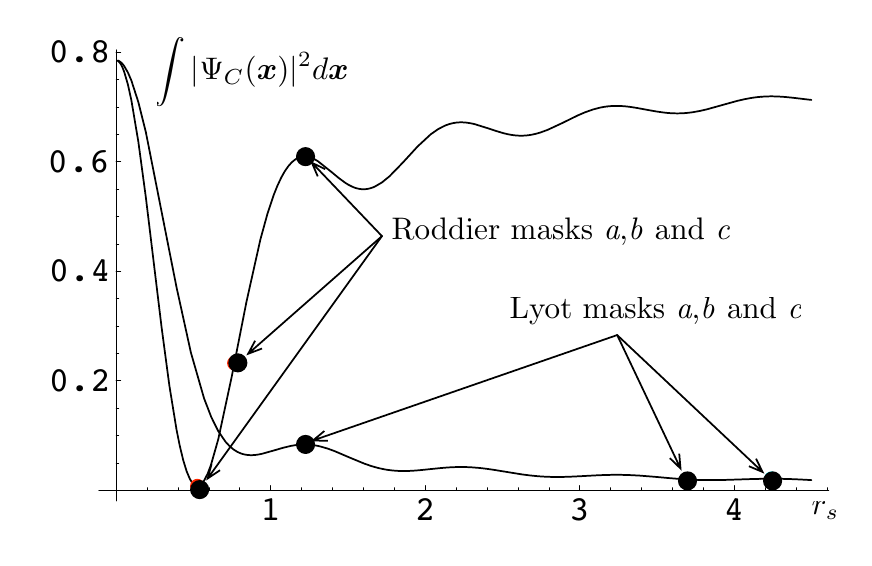}}
\caption{The left plot locates the different masks used in the simulation on 
the response of an on-axis point source. The right plot gives the integrated intensity in $C$ for $r_s=R$ as a function of the mask size. 
\label{fig:taillmask}}
\end{figure}

Figure \ref{fig:lyot} illustrates the response of the Lyot coronagraph to an on-axis point source 
($\Psi(\vec{x})=1$) for different values of the mask radius $r_m$. 
These values of $r_m$  are located on $\hat{\Psi}_A(r)$ in figure \ref{fig:taillmask}.

We seek to obtain the best subtraction of the two wavefronts $\Psi_A(r)$ and
$\Psi_A(r)\ast \hat{m}(r)$ inside de Lyot stop. Figure \ref{fig:lyot} cleary shows that this result is achieved
increasing the mask size.
In fact, as $r_m$ increases, $\hat{m}(\vec{x})$ will be more ``concentrated''
around the origin. As long as  $ \hat{m}(\vec{x})$ always verifies $\int \hat{m}(\vec{x})d\vec{x}=m(\vec{0})=1$ 
we have $\hat{m}(\vec{x}) \rightarrow \delta(\vec{x})$ as $r_m \rightarrow +\infty$, which is
the neutral element for the convolution product. On the contrary as    $r_m$ decreases  $\hat{m}(r)$
widens around $\hat{m}(0) =\pi r_m^2$. At the limit $r_m \rightarrow 0$,
$\hat{m}(\vec{x}) \rightarrow \pi r_m^2$  and $\Psi_A(\vec{x})\ast \hat{m}(\vec{x})$ tends to the constant $\pi^2 R^2 r_m^2$, which does not substract to $\Psi_A(\vec{x})=1$. It is interesting to note that this problem
is formally equivalent to the problem of digital low-pass filter design with
finite impulse response using the windowing method where the infinite
impulse response of the ideal filter is truncated using a window, see for example \citep{Oppenheim1989dts}.

The results of figure \ref{fig:lyot} suggest the following comments:
\begin{itemize}
\item The performance of the coronagraph is not strictly linear with the mask size. The choice
of mask \textit{b} is as an evidence more appropriate than mask \textit{c}.
This is confirmed by the general shape of the  second plot in figure \ref{fig:taillmask} which gives the integrated
intensity in $C$ for a Lyot stop of radius $R$.

\item A large amount of the residual star energy is located at the edges of the pupil. Consequently
a moderate reduction of the diameter of the Lyot stop radius $r_s$ will provide  a  significant gain
in starlight rejection with a reasonable loss of transmission. The effect of a Lyot stop reduction is
visible in the last two lines of the plot. The value of $r_s$ for the last line is indicated in
the plots of the intensity in $C$ by a dashed line.

\item The off-axis transmission of the coronagraph is of course a crucial point for planet detection.
As long as the angular distance from the optical axis is sufficient to guaranty that the response
of the planet in $B$ will not interfere with the mask, we can consider that the response of the 
planet will be the shifted PSF of a pupil with radius $r_s$. In order to quantify the loss of transmission
for an off-axis source, this PSF has been added in figure \ref{fig:lyot}.
Finally, note that a solution to the analytical computation of $\Psi_C(r,\theta)$, $\Psi_D(r,\theta)$ 
of a point source which can be close to the optical axis has been addressed in  \citep{art_apj07}. 

\end{itemize}

\begin{figure}[ht]
\centerline{\includegraphics[width=.8\textwidth]{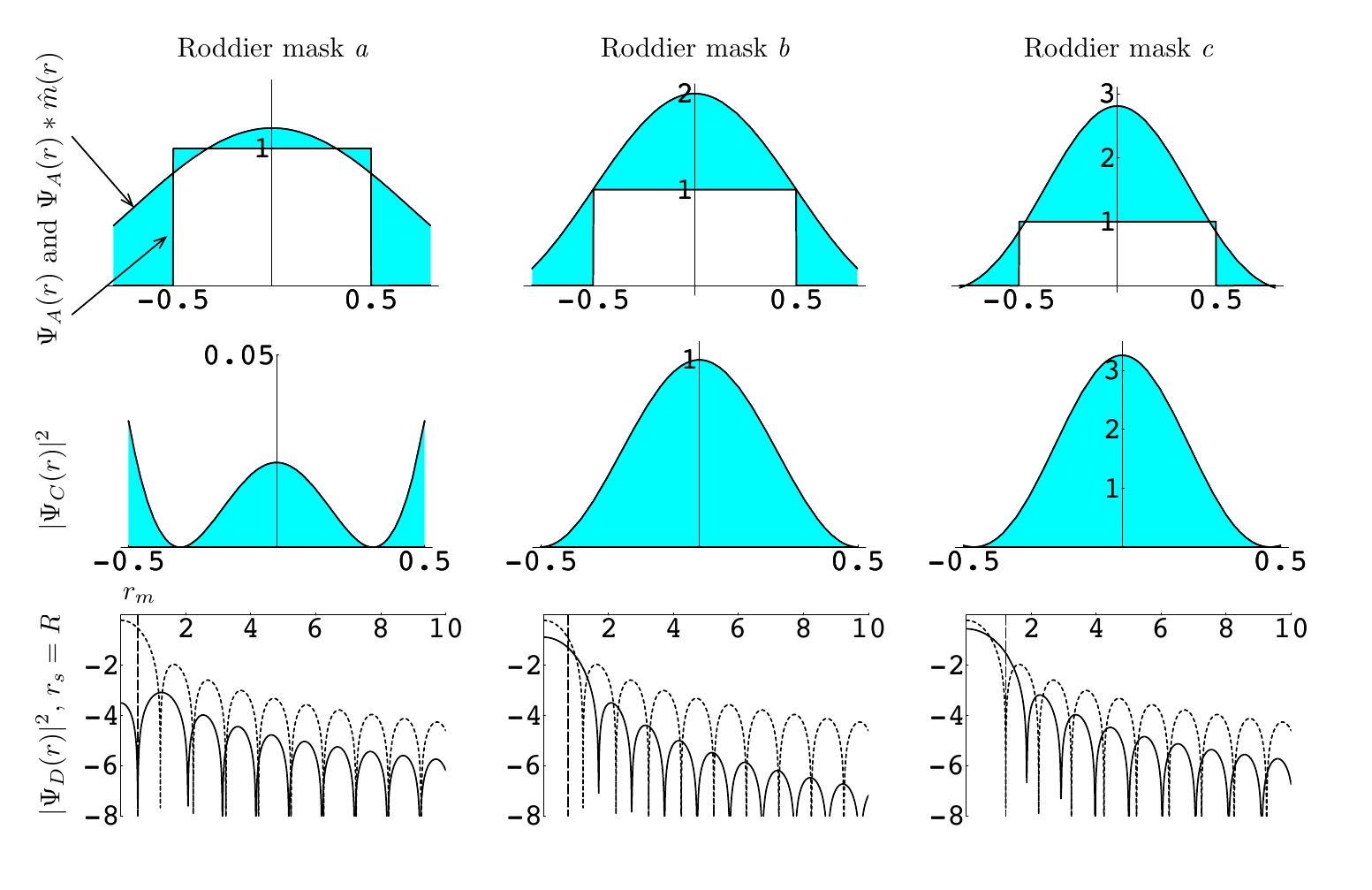}}
\caption{Illustration of Roddier coronagraphy for circular aperture (radial cuts) and an on-axis  point source. 
The first raw represents the two complex amplitudes 
that should interfere destructively in $C$.
The second row gives the intensity in $C$ and the third 
raw the intensity in $D$.
In the last row, the continuous line represents the intensity in $D$ for a coronagraphed
point source and the dashed line the PSF associated to a pupil of radius $r_s$.
\label{fig:roddier}}
\end{figure}

The Lyot technique was improved by Roddier  \citep{rodd97} replacing the opaque mask by a $\pi$ phase mask: $m(\vec{x}) = 2\Pi(r/r_m)$.
Figure \ref{fig:roddier} show that for very small values of $r_m$, the Roddier coronagraph tends to behave like
the Lyot coronagraph, trying to cancel out the light of the star. This behavior totally
differs when $r_m$ increases: for mask size \textit{a} or \textit{b} the effect of this coronagraph is to produce in the plane \textit{C} an apodized version 
of the wavefront originated by an on-axis source.
Note that a coronagraphic technique relying on a similar principle was proposed in  \cite{martina04} using
a $\pi/2$ phase mask and a defocus.
The effect of this apodization is visible in \textit{D}: the main lobe of the response is broadened whereas the side lobes decrease. Moreover, it is worthy to
note that this apodization as a maximum value for $\vec{x}=0$ that can be
higher than 1. 

These results suggest the following comments:
\begin{itemize}
\item the choice of $r_m$ remains a crucial point. However, according to
 the preceding remark  the integrated energy in \textit{C} ploted in figure
\ref{fig:taillmask}  is no more a valid criterion
 for selection of an optimal value of $r_m$. 
 
 \item As the second row of  figure \ref{fig:taillmask}  shows, a reduction 
 of the Lyot stop is not crucial as long as the  Roddier coronagraph operates
 in its `` apodization mode''.

\item The effect of a Roddier coronagraph on a planet relies on the same principle
as the Lyot coronagraph. Consequently the apodization effect mentioned above will not operate
for an off-axis source. Moreover, the fact that for this latter coronagraph  the mask
will be smaller and a reduced stop is not necessary will increase
 its performances in terms of resolution and planet flux. 

\end{itemize}
According to the previous remarks,  a Roddier coronagraph with an extended mask
can be used as an apodizer for the  wavefront complex amplitude.  
The star rejection of such a system being insufficient for planet detection such a
device should be considered as the first stage of  a ``classical'' coronagraph.

\section{Optimal coronagraphy}

According to equation (\ref{planCcart}) the ideal coronagraph equation is:
\begin{equation}
\Psi_A(\vec{x})  = \Psi_A(\vec{x}) \ast \hat{m}(\vec{x}),\; \forall \vec{x}\in \mathcal{S}
\label{eqvp}
\end{equation}
where $\Psi_A(\vec{x}) = a(\vec{x})p(\vec{x})$.
Following Lyot and Roddier coronagraphs, a first solution is to set
$m(r)=\epsilon \Pi(r/r_m)$ where $\epsilon=1$ for Lyot and
$\epsilon=2$ for Roddier and to find, if it exists,
the optimal apodization $a(\vec{x})$ solution of the integral equation (\ref{eqvp}). This solution was first proposed by \citep{baudth} and 
\citep{guyon2000}.
The other solution is to suppress the constraint on the mask and to find
the function $m(\vec{x})$ solution of equation (\ref{eqvp}) when 
$a(\vec{x})=1$. This approach includes the band-limited masks
proposed by  \citep{kuch02} and the four quadrants coronagraph 
\citep{roua00}.

\subsection{Finite size masks: apodized Lyot and Roddier coronagraphs}

As long as $m(\vec{x})$ has a bounded symmetric support $\mathcal{M}$, we can write 
using the notations of Apendix B: $\hat{m}(\vec{x}) = \epsilon K_M(\vec{x})$. 
Consequently, equation (\ref{eqint1}) implies that
if $a(\vec{x})$ is proportional to a spheroidal prolate 
function associated to $\mathcal{M}$ and $\mathcal{P}$, the residual in $C$ when $p_s(\vec{x})=p(\vec{x})$ is obtained
replacing in (\ref{planCcart}) the second term by $\epsilon\Lambda \Psi_A(\vec{x})$:
\begin{equation}
\Psi_C(\vec{x}) = (1-\epsilon\Lambda)\Psi_A(\vec{x}) \label{repprolenC}
\end{equation}
which states that the residual wavefront in $C$  originated by an on-axis source is
proportional to the apodized  entrance pupil. If we assume that the angular distance
of the planet is sufficient, its response in $C$ is obtained setting  $\epsilon=0$ in (\ref{repprolenC}).
The tradeoff between the reduction of the planet flux by the apodization and the reduction of
the star flux by the coronagraph can be evaluated by the quantity $\eta$ defined as the ratio
between the total planet intensity in $D$ and the  total star intensity outside the mask in $D$.

From equation (\ref{repprolenC}) the star intensity in $D$ is simply $(1-\epsilon\Lambda)^2|\hat{\Psi}_A(\vec{x})|^2$. The total energy outside the mask is:
\begin{equation}
(1-\epsilon\Lambda)^2\big( \int_{\mathbb{R}^2}|\hat{\Psi}_A(\vec{x})|^2d\vec{x}
- \int_{\mathcal{M}}|\hat{\Psi}_A(\vec{x})|^2d\vec{x}  \big)
=(1-\epsilon \Lambda)^2\big(\int_{\mathbb{R}^2}|\hat{\Psi}_A(\vec{x})|^2d\vec{x}
-\Lambda \int_{\mathbb{R}^2}|\hat{\Psi}_A(\vec{x})|^2d\vec{x} \big) 
\end{equation}
where the last equality comes from the fact that $\epsilon\Lambda$ is the ratio between the energy 
encircled in the  bounded frequency domain $\mathcal{M}$ and the total energy of $\Psi_A(\vec{x})$.
As a consequence we have:
\begin{equation}
\eta = (1-\epsilon \Lambda)^2(1-\Lambda) \label{eta}
\end{equation}
The objective will be to minimize $\eta$ for a given coronagraph type ($\epsilon=1$ or 2). 
This is achieved by a modification of the size of $\mathcal{M}$ which will
affect $\Lambda$ and the pupil apodization shape $a(\vec{x})$.

Any eigenvalue of the integral equation minimizing (\ref{eta}) 
leads to a valid solution. However we will only consider in the sequel
the  largest eigenvalue, i.e. corresponding to the maximum encircled 
energy behind the focal plane mask.
This choice will in fact ``generally'' lead to a positive prolate function
which will be normalized by its maximum value on the pupil
in order to achieve an apodization with maximum transmission.
Finally, it is worthy to note that this development does not require a specific shape of the pupil and the mask. 
However, further analytical derivations based on equations (\ref{eqint2}) and (\ref{eqint3}) require that
the mask is a scaled version of the pupil: in this case if the pupil has a central obstruction 
$m(\vec{x})$ has a hole. This point will not be considered herein.

\begin{figure}[ht]
\centerline{\includegraphics[width=.7\textwidth]{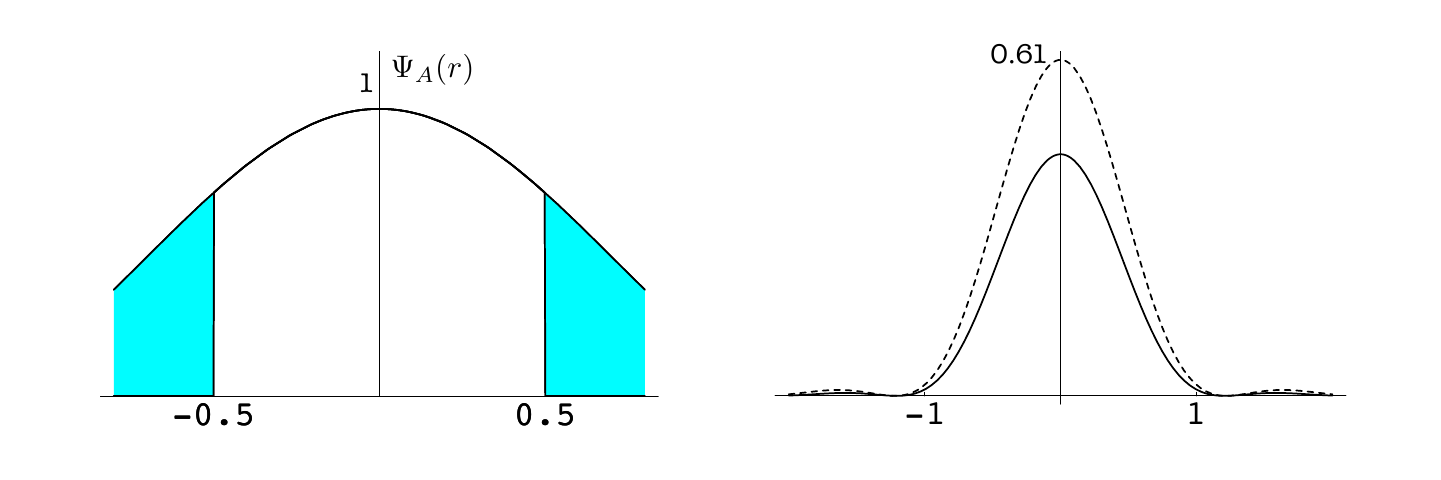}}
\caption{Illustration of Roddier apodized coronagraphy for circular aperture (radial cuts). The left plot shows the optimally prolate apodized  pupil.
The right plot shows the PSF corresponding to the prolate apodized pupil
and the PSF of the unapodized pupil. \label{fig:roddierapo}}
\end{figure}

\begin{figure}[ht]
\centerline{\includegraphics[width=.8\textwidth]{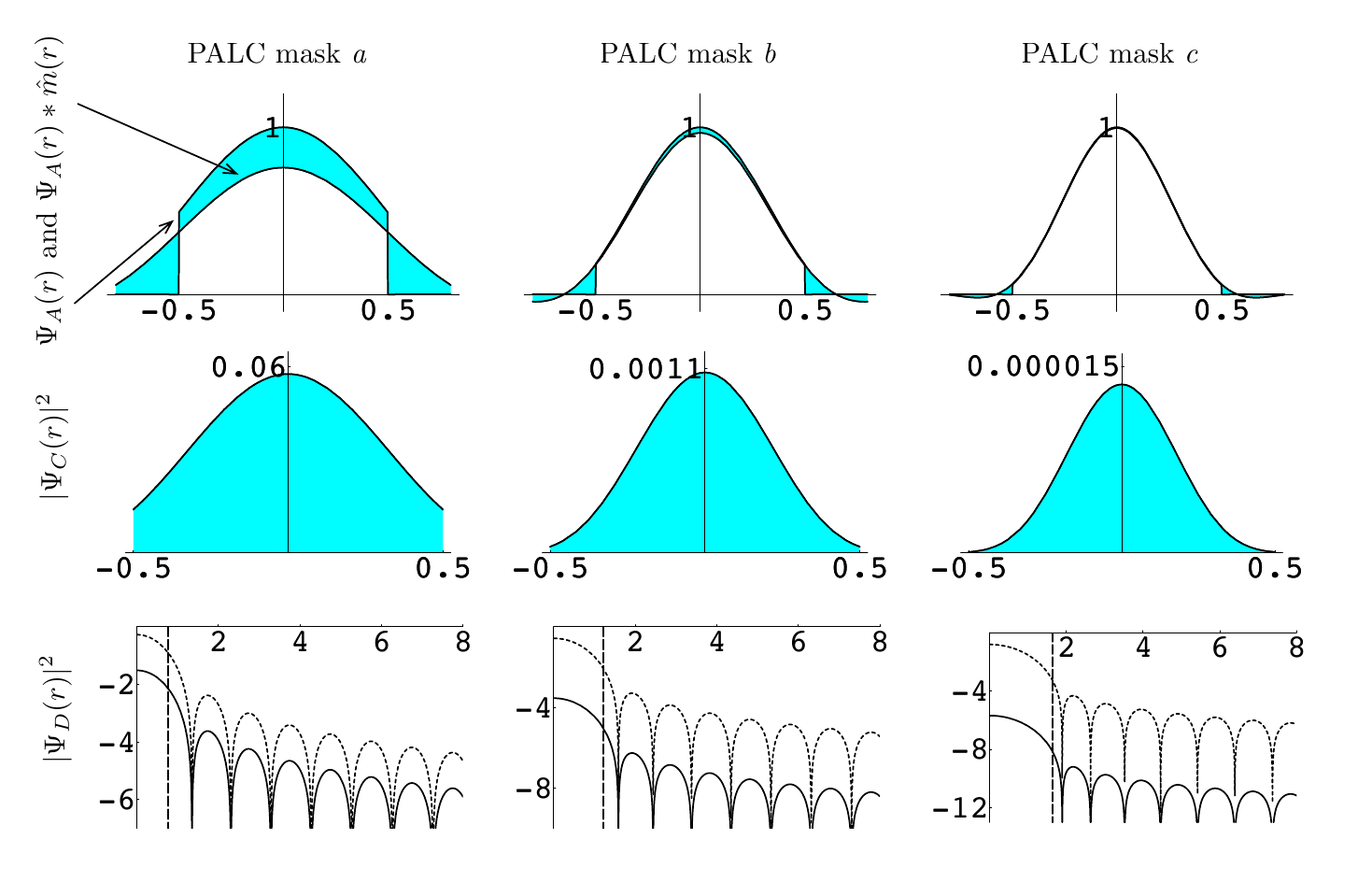}}
\caption{Illustration of prolate apodized coronagraphy for circular aperture (radial cuts) and an on-axis  point source. 
The first raw represents the two complex amplitudes 
that should interfere destructively in $C$.
The second row gives the intensity in $C$ and the third and fourth
raw the intensity in $D$.
In the last row, the continuous line represents the intensity in $D$ for a coronagraphed
point source and the dashed line the PSF associated to the apodized telescope
pupil.
\label{fig:lyotapo}}
\end{figure}

\begin{itemize}
\item We consider first the case of an apodized Roddier coronagraph, $\eta = (1-2\Lambda)^2(1-\Lambda)$.
The results of Appendix A prove that 1/2 is a valid eigenvalue and consequently
a mask size can be chosen to achieve $\eta=0$, i.e. a total extinction of the star. 

For example, in the case of a circular pupil without obstruction,
figure \ref{fig:desprol} shows that the coefficient $c=r_mR\approx 0.25$ 
corresponds to $\Lambda = 1/2$. This value defines
the  apodization shape and fixes the mask size to $r_m\approx 0.25/R$. 
Figure \ref{fig:roddierapo} gives the corresponding apodization.
This figure also illustrates the loss in transmission for an off axis planet  
comparing  the prolate apodized pupil PSF with the PSF of the unapodized pupil.

\item We consider now a prolate apodized Lyot coronagraph (PALC), $\eta = (1-\Lambda)^3$.
The eigenvalues being upper-bounded by 1, a prolate apodized Lyot coronagraph (PALC) cannot
achieve total extinction of an on-axis point source. The trivial solution $\Lambda= 1$
would correspond to an infinite size opaque mask.
However, approximate solutions can be obtained for eigenvalues $\Lambda$ close to 1 
and finite mask size. Taking advantage of the rapid saturation of the eigenvalue curve
we can choose a mask size  corresponding to an eigenvalue close to 1.

Figure \ref{fig:lyotapo} illustrates the response of the PALC to an on-axis point source 
for different values of the mask radius $r_m$ given in figure \ref{fig:taillmask}. 
To each value of $r_m$ corresponds a coefficient $c$ and and
optimal apodization function given in the first row. The second row gives the residue
in $C$ and the last row the response of the star and the PSF associated to the apodized telescope pupil. Note that contrarily to the unapodized Roddier coronagraph operating in its ``apodized regime'' (see column 2 and 3
of figure \ref{fig:roddier}) the residue in $C$ is here both apodized \emph{and}
attenuated. An analog comment can be done for the apodization alone
coronagraphs.

The property that the Lyot stop wave amplitude is proportional to the entrance apodized pupil amplitude, creates the possibility for multiple stage coronagraphs described in \cite{aime04}.
A multiple stage PALC only requires a single apodizer in the entrance pupil.  The Lyot stop plane is naturally apodized and can be used as the entrance pupil of a second coronagraphic stage
 without further loss of throughput due to apodization. In this case the residual
 amplitude in the second Lyot stop plane is $(1-\Lambda)^2\Psi_A(\vec{x})$
 and the parameter $\eta$ becomes $(1-\Lambda)^5$.
\end{itemize}

\subsection{Infinite size masks band limited and  four quadrants coronagraphs}

\begin{figure}[ht]
\centerline{\includegraphics[width=.7\textwidth]{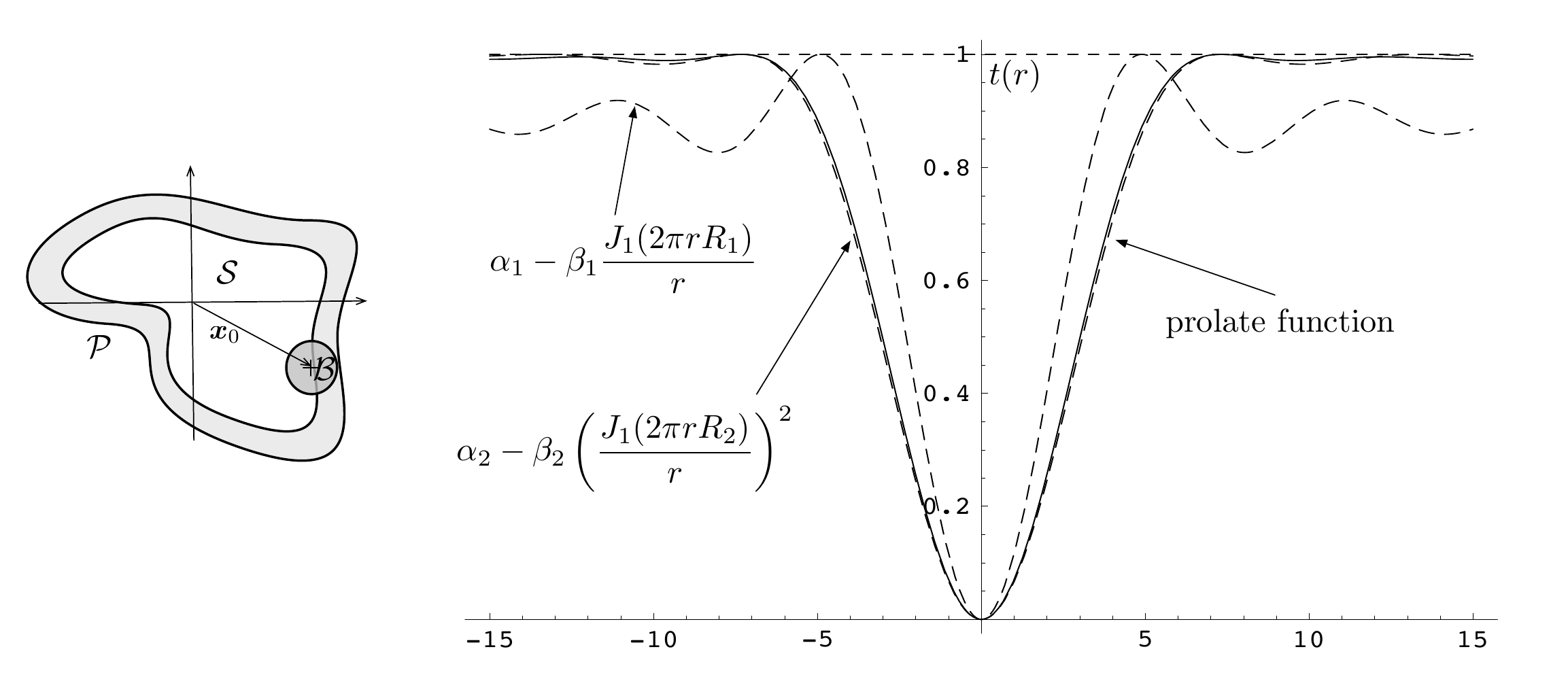}}
\caption{Left figure: relation between the pupil ($\mathcal{P}$), support of
the mask Fourier transform ($\mathcal{B}$) and the Lyot stop $\mathcal{S}$
in a bandlimited coronagraph. Right figure: illustration of three different masks.
Dashed lines: second and fourth order Bessel cardinal masks. Continuous line:
prolate spheroidal function. \label{fig:BS}}
\end{figure}

A simple solution to define a mask $m(\vec{x})$ that verifies
 (\ref{eqvp}) is to chose $m(\vec{x})$ such that
$\hat{m}(\vec{x})$ is non zero only on a small bounded region $\mathcal{B}$
included inside the pupil. In this case if we define $\mathcal{S}$ such as 
$\forall \vec{x}_0 \in \mathcal{S}$   the support of $\hat{m}(\vec{x}_0-\vec{x})$ is strictly included in the pupil $\mathcal{P}$:
\begin{equation}
\forall \vec{x}_0 \in \mathcal{S},\; \Psi_A(\vec{\vec{x}_0})\ast \hat{m}(\vec{x}_0) = \int_\mathcal{P} \hat{m}(\vec{x}_0-\vec{x})d\vec{x}=m(\vec{0})
\end{equation}
Consequently, if we impose $m(\vec{0})=1$ we will clearly
verify equation (\ref{eqvp}) on $\mathcal{S}\subset\mathcal{P}$.
Figure \ref{fig:BS} illustrates the definition of the various sets $\mathcal{S}$, $\mathcal{P}$ and $\mathcal{B}$. 
A simple solution to block the residue in $\mathcal{P}\backslash\mathcal{S}$
is to use as a Lyot stop the indicator function of $\mathcal{S}$, 
$p_s(\vec{x}) = \nbOne_\mathcal{S}(\vec{x})$ ($p_s(\vec{x})=1$ if $\vec{x}\in \mathcal{S}$ and $0$ elsewhere).
Consequently a properly defined band limited, and necessary 
infinite size mask $m(\vec{x})$ must verify:
$m(\vec{0})=1$, $0\leq m(\vec{x})\leq 1$ and
$m(\vec{x})$ is bandlimited on $\mathcal{B} \subset \mathcal{P}$.
 This approach was proposed by \citep{kuch05}.

A fundamental result for understanding of band limited coronagraphs states that
the complex amplitude in $D$ associated to an off-axis point source at an angle 
$\vec{x}_1$ is the response of the Lyot stop attenuated by the transmission at
$\vec{x}_1$:
\begin{equation}
\Psi_D(\vec{x}) = t(\vec{x}_1)\hat{p}_s(\vec{x}-\vec{x}_1) \label{eqkuch}
\end{equation}
This result is demonstrated in \citep{kuch05}. This result suggests the following
 comments:
 \begin{enumerate}
 \item The loss of throughput of a bandlimited coronagraph is principally due
 to the undersizing of the Lyot stop. Consequently $\mathcal{B}$ should be as small
 as possible. However the general consequences of a reduction of $\mathcal{B}$ are
 a widening of the central lobe of $m(\vec{x})$ and an increase of the ripples in the tails of
  $m(\vec{x})$. From equation (\ref{eqkuch})   the first effect implies
  an increase of a blind zone arround the star and the second  a reduction of detectability  
  for given locations of a planet.
  
  \item   The behavior of the coronagraph when the star is slightly off-axis
 is perfectly described by equation (\ref{eqkuch}): in order to reduce the effect of a misalignmemt
   $m(\vec{x})$ must be as ``flat'' as possible  in $\vec{0}$.   
    This effect is quantified in \cite{kuch05}   by the degree of 
 the first term in the serie expansion of $t(\vec{x})$.
  Note that the flatness of $m(\vec{x})$ around $\vec{x}=\vec{0}$ is of course  deeply related to the constraints previously mentioned.
Finally, a technique  to construct a  mask of given order adding and multiplying simple  band-limited masks is proposed in \cite{kuch05}.

 \end{enumerate}

To conclude, figure \ref{fig:BS} compares the  behavior of three different radial transmissions in the case of a circular aperture. To facilitate this comparison  the three masks have been
normalized to the same $\mathcal{B}$ in order to guaranty that the Lyot stop will be the same:
for these simulations the flux reduction equals $4/9$. The first  transmission
is a Bessel cardinal function and the second one the square of a  Bessel cardinal function. These functions have been properly scaled and shifted in amplitude in order that $t(r)$ occupies all the interval $[0,1]$.
The properties detailed in the  item (i) of the  discussion above can also be
expressed in term of concentration of the energy of $m(\vec{x})$ around
the origin. In this case, the additional band-limited constraints leads naturally 
to a prolate function. Such a solution is illustrated in figure \ref{fig:BS}.
In this case the coefficient $c$ has been chosen empirically equal to 6.8 
in order to achieve a good compromise between the the ripples and the width
of the main lobe. This result shows that the behavior of such a prolate mask and
a squared Bessel cardinal mask are similar. Finally, it is important to emphasize
that the masks illustrated in  figure \ref{fig:BS} correctly  behave
with respect to the requirement (i).  On the contrary they do not fulfil (ii)
(the order of the squared Bessel cardinal and the prolate is only 4) 
and will have a poor behavior in the case of a resolved star or a misalignment.

To conclude, it is important to mention in this section devoted to optimal coronagraphy with infinite size masks the 4QC. This coronagraph will be developed in a
separate paper by D. Rouan in this volume. 
It relies on the fact that for a circular aperture and  a transmission  
$t(\vec{x})=\text{sign}(x)\text{sign}(y)$, the complex amplitude  in $C$ is identically zero inside the pupill for a point source on the axis. 
This beautiful result relies on a nontrivial property of the Fourier transform (see the
paper of D. Rouan and included references for a proof). 
A coronagraphic technique based   on a similar principle is developped in \cite{vortex05}.

\section*{Acknowledgements}
The authors would like to thank Peter  Falloon for his Mathematica
program.

\appendix

\section{Prolate spheroidals functions}

This section presents some facts about  prolate spheroidals functions.
For a detailed presentation  refer to the seminal papers
\citep{slep64} and references therein. Application of prolate spheroidals functions to optics
can be found in \citep{friedb}.  

In order to gain deeper insight in their principal  properties using simple mathematics the cartesian coordinates have been prefered. 
Derivations using specific coordinates can be found in the references.
We consider a real valued square-integrable function $f$ of $d$ variables having a  bounded
support  $\mathcal{P}$ (in our case $f(\vec{x})=a(\vec{x})p(\vec{x})$
and $\mathcal{P}$ is the telescope pupil):
 \begin{equation}
\hat{f}(\vec{x}) =  \int_\mathcal{P} f(\vec{y})e^{-\imath 2 \pi 
\vec{x}^t\vec{y}}d\vec{y} \label{eq:bft}
 \end{equation}

 The ratio between the energy encircled in the  bounded frequency domain $\mathcal{M}$
 (here the Lyot mask) and the total energy of $\hat{f}$ is obtained integrating the
 squared modulus of (\ref{eq:bft}) on $\mathcal{M}$:
 \begin{equation}
 \int_\mathcal{P} K_\mathcal{M}(\vec{x}-\vec{y})f(\vec{y})f(\vec{x})
 d\vec{y}d\vec{x}\big/ \int_\mathcal{P} f(\vec{y})^2 d\vec{y}\label{ratio1}
 \end{equation}
 where $K_\mathcal{M}(\vec{x})$ is the inverse Fourier transform 
of the indicator function of $\mathcal{M}$, $\nbOne_\mathcal{M}(\vec{x})$. Standard results on functional analysis, see for example \citep{rieszb}, prove that the maximum of this ratio
is the largest eigenvalue $\Lambda$ of the integral equation:
\begin{equation}
\int_\mathcal{P} K_\mathcal{M} (\vec{x}-\vec{y}) \psi(\vec{y})d\vec{y}=\Lambda \psi(\vec{x})
,\; \vec{x}\in \mathcal{P} \label{eqint1}
\end{equation}
or equivalently:
$\big(\psi(\vec{x})
\nbOne_\mathcal{P}(\vec{x})\big)
\ast K_\mathcal{M} (\vec{x})
=\Lambda \psi(\vec{x})$, $\vec{x}\in \mathcal{P}$.

The prolate spheroidal functions are defined as the solution of this integral equation.
The extrema of  (\ref{ratio1}) is reached when $f(\vec{x})$ is
the corresponding eigenfunction. The kernel being positive defined the eigenvalues of
(\ref{eqint1}) are positive and from (\ref{ratio1}) upper bounded by 1. The eigenfunctions of (\ref{eqint1})  are orthogonal and complete on 
$L_2(\mathcal{P})$ and orthogonal  on $\mathbb{R}^d$ when extended outside $\mathcal{P}$ using  (\ref{eqint1}). 

Considerable simplifications occur  when $\mathcal{P}$ and $\mathcal{M}$ are both
scaled versions of a ``normalized'' domain  $\mathcal{U}$:
$\exists \vec{R}\in \mathbb{R}^d$, $\exists \vec{r}_m\in \mathbb{R}^d$ such that $\vec{x}\in \mathcal{P}  \Leftrightarrow \vec{x}./\vec{R} \in \mathcal{U}$,  $\vec{x}\in \mathcal{M}  \Leftrightarrow \vec{x}./\vec{r}_m \in \mathcal{U}$,
where $\vec{a}./\vec{b}$ (resp. $\vec{a}.\vec{b}$) defines the vector having $a(k)/b(k)$ (resp. $a(k)b(k)$) as components.
A simple change of variable in the definition of $K_\mathcal{M}(\vec{x})$ shows that
$K_\mathcal{M}(\vec{x}) = \left(\prod_{k=1}^d r_m(k)\right) K_\mathcal{U}(r_m.\vec{x})$. Substitution of
this result in  (\ref{eqint1}) shows that $\psi(\vec{x})=\phi(R.\vec{x})$ where
$\phi(\vec{x})$ is the solution of the normalized problem:
\begin{equation}
 \left(\prod_{k=1}^d r_m(k)R(k)\right) \int_\mathcal{U} K_\mathcal{U} (\vec{r}_m.\vec{R}.(\vec{x}-\vec{y})) \phi(\vec{y})d\vec{y}=\Lambda 
\phi(\vec{x})
,\; \vec{x}\in \mathcal{U} \label{eqint2}
\end{equation}
\begin{itemize}
\item Consider for example the case where $d=2$ and $\mathcal{U}$ is a disk. Consequently $ \mathcal{P}$ and $\mathcal{M}$ are ellipses. Equation (\ref{eqint2}) shows that  the prolate functions associated to an ellipse 
are obtained by a scaling of the  prolate function associate to a disk.

\item If $\forall k,\; R(k)=R$ and $r_m(k)=r_m$, equation (\ref{eqint2}) shows that  the prolate spheroidal  functions for a given $\mathcal{U}$ form essentially a one-parameter family of functions depending of the parameter $c=r_mR$.
\end{itemize}

Additional simplifications in this last case are obtained when the region $\mathcal{U}$ is symmetric:
$\vec{x}\in \mathcal{U} \Leftrightarrow  -\vec{x}\in \mathcal{U}$. In this case
the solutions of (\ref{eqint2}) are real, either even with a real eigenvalue or odd
with a pure imaginary eigenvalue. 
Moreover, finding the
solutions of (\ref{eqint2})  is equivalent to finding the solutions of:
\begin{equation}
\alpha  \phi(\vec{x})  = \int_\mathcal{U} e^{-\imath 2\pi c  \vec{x}^t\vec{y}}
\phi(\vec{y})d\vec{y},\; \vec{x}\in \mathcal{U},\;
\text{where } \Lambda = |\alpha|^2 c^d
\label{eqint3}
\end{equation}
This equation states that the Fourier transform of $\phi(\vec{x}) \nbOne_\mathcal{U}(\vec{x})$
is proportional to $\phi(\vec{x}/c)$.
This result can be easily verified taking the complex conjugate of
(\ref{eqint3}), substituting in the integral $\phi(\vec{y})$ by the right side of
(\ref{eqint3}) and identifying with (\ref{eqint2}).

\begin{figure}[ht]
\centerline{\includegraphics[width=.8\textwidth]{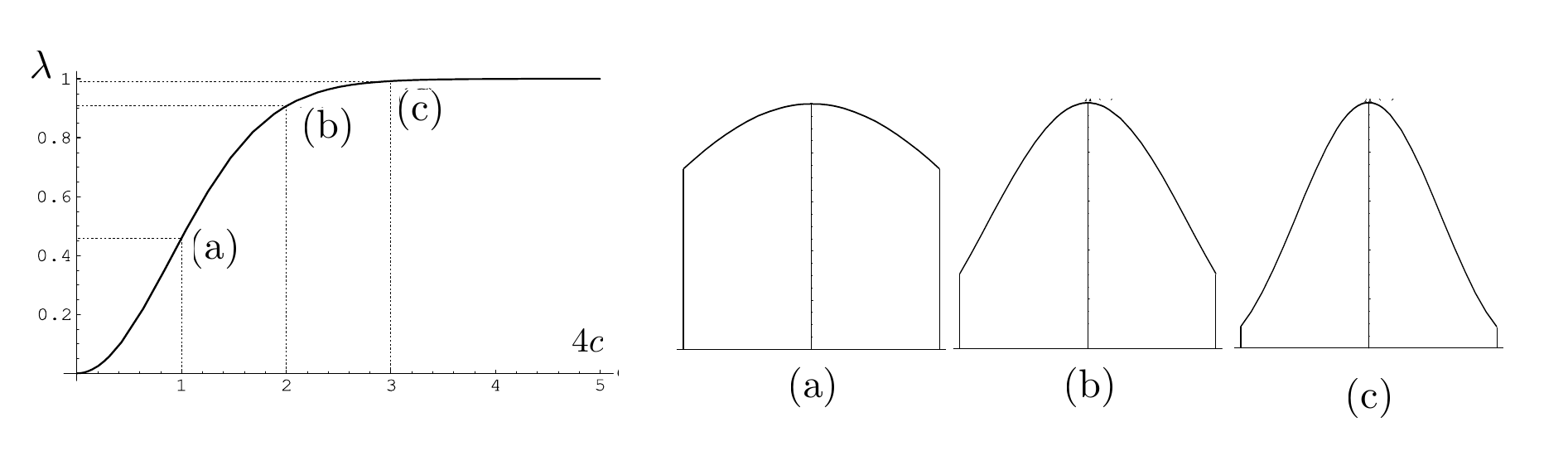}}
\caption{Circular prolate functions. Left plot gives the largest eigenvalue $\Lambda$ as a function of $4c$. The right plot gives the prolate functions for three values of $c$.\label{fig:desprol}}
\end{figure}

Circular prolate functions correspond to the case $d=2$
where $\mathcal{U}$ is the unit disk.  
They decompose, up to a normalisation factor  as $\phi_{N,n}(r,\theta)=R_{N,n}(r)[\cos,\sin](N\theta)$.  Consequently (\ref{eqint3}) shows that 
they can be defined by their invariance to a finite Hankel transform of order $N$,
e.g.:
\begin{equation}
2\pi \int_0^1 R_{0,0}(r)J_0(2\pi r\rho)rdr  = \frac{\sqrt{\Lambda}}{r_m }
R_{0,0}\left(\frac{\rho}{r_m}\right)
\end{equation}
Figure \ref{fig:desprol} represents the radial circular prolate function 
$\phi_{0,0}(r)$ for three different values of $c$.

A crucial point is the numerical computation of prolate functions. For simple geometries, the functions can be computed using rapidly converging
series as the one derived for a circular aperture in \citep{slep64}. For general
geometries, prolate spheroidal functions can be computed directly 
solving equation (\ref{eqint1}) using one of the numerous iterative algorithm
that can be found in the litterature for the computation of the eigenelements 
of linear integral operators, e.g. the algorithm used in \citep{guyon2000}.

%\bibliographystyle{elsart-harv}
%\bibliography{../BIBLIO/labibli}
  
\end{document}